\documentclass[prx,aps, nobalancelastpage,superscriptaddress,nolongbibliography, onecolumn]{revtex4-2}

\usepackage{color,amsthm,amsmath,amsxtra,amsfonts,dsfont,graphicx,bm,amssymb}
\usepackage[colorlinks=true,linkcolor=blue, citecolor=blue, urlcolor=blue, bookmarks]{hyperref}
\usepackage{centernot}
\usepackage[dvipsnames]{xcolor}
\usepackage{graphicx}
\usepackage{tikz}
\usepackage{braket}
\usepackage{multirow}
\usepackage{comment}
\usepackage{makecell}

\newcommand{\be}{\begin{equation}}
	\newcommand{\ee}{\end{equation}}
\newcommand{\bea}{\begin{eqnarray}}
	\newcommand{\eea}{\end{eqnarray}} 
\newcommand{\bse}{\begin{subequations}}
	\newcommand{\ese}{\end{subequations}}

\theoremstyle{plain}

\newcommand{\tr}{\mathrm{Tr}}

\newcommand{\prlsection}[1]{{\em {#1}.---~}}

\usepackage{pifont}
\usepackage{tikz}
\usetikzlibrary{decorations.pathreplacing,calligraphy,decorations.markings}

\begin{document}
	\date{\today}

	\newcommand{\bbra}[1]{\<\< #1 \right|\right.}
	\newcommand{\kket}[1]{\left.\left| #1 \>\>}
	\newcommand{\bbrakket}[1]{\< \Braket{#1} \>}
	\newcommand{\pll}{\parallel}
	\newcommand{\nn}{\nonumber}
	\newcommand{\transp}{\text{transp.}}
	\newcommand{\nor}{z_{J,H}}
	
	\newcommand{\hL}{\hat{L}}
	\newcommand{\hR}{\hat{R}}
	\newcommand{\hQ}{\hat{Q}}

\title{An asymmetry lower bound on fermionic non-Gaussianity}

\begin{abstract}
Fermionic Gaussian states are a fundamental tool in many-body physics, faithfully representing non-interacting quantum systems and allowing for efficient numerical simulations. Given a many-body wave function, it is therefore interesting to ask how much it differs from that of a Gaussian state, as quantified by the notion of non-Gaussianity. In this work, we relate measures of non-Gaussianity with the Shannon entropy of the particle-number distribution, coinciding with the particle-number asymmetry for pure states. We derive a lower bound on the relative entropy of non-Gaussianity in terms of the exponential of the Shannon entropy, and study numerically its tightness for large system sizes. Our bound is non-trivial for large values of the asymmetry and relies on the concentration of the particle-number distribution of (mixed) fermionic Gaussian states. Since the Shannon entropy of the particle-number distribution is often efficient to compute or experimentally measure, our results can be viewed as a practical way to lower bound non-Gaussianity,  highlighting a non-trivial interplay with particle-number asymmetry.

\end{abstract}

\author{Filiberto Ares}
\affiliation{SISSA and INFN, via Bonomea 265, 34136 Trieste, Italy}
\author{Michele Mazzoni}
\affiliation{Dipartimento di Fisica e Astronomia, Universit\`a di Bologna and INFN, Sezione di Bologna, via Irnerio 46, 40126 Bologna, Italy}
\author{Sara Murciano}
\affiliation{Universit\`e Paris-Saclay, CNRS, LPTMS, 91405, Orsay, France}
\author{D\'avid Sz\'asz-Schagrin}
\affiliation{Dipartimento di Fisica e Astronomia, Universit\`a di Bologna and INFN, Sezione di Bologna, via Irnerio 46, 40126 Bologna, Italy}
\author{Pasquale Calabrese}
\affiliation{SISSA and INFN, via Bonomea 265, 34136 Trieste, Italy}

\author{Lorenzo Piroli}
\affiliation{Dipartimento di Fisica e Astronomia, Universit\`a di Bologna and INFN, Sezione di Bologna, via Irnerio 46, 40126 Bologna, Italy}

\maketitle
	

\section{Introduction}

Fermionic Gaussian states are a very useful tool in condensed matter and many-body physics~\cite{bravyi2004lagrangian, surace2022tagliacozzo}. Besides faithfully representing the eigenstates of non-interacting fermionic Hamiltonians~\cite{franchini2017introduction}, their simplified mathematical structure makes them ideal toy models that are routinely employed in different areas of physics. For instance,  in the context of quantum computation, fermionic Gaussian states naturally appear using the Jordan-Wigner (JW) transformation~\cite{jordan1993}, mapping spin (or qubit) degrees of freedom to fermionic ones. The circuits made of gates that can be mapped to free-fermion dynamics in this way are known as matchgate quantum circuits~\cite{terhal2002classical}. They play an important role, since the mapping to free fermions makes it possible to simulate them efficiently on a classical computer~\cite{vanDenNest2011simulating,brod2016efficient}. More generally, fermionic Gaussian states are becoming increasingly relevant in the development of new quantum technologies. For example, several algorithms now allow us to efficiently learn an experimentally prepared fermionic Gaussian state~\cite{gluza2018fidelity,aaronson2021efficient,ogorman2022fermionic,mele2025efficient,bittel2025optimal}, offering non-trivial benchmarking opportunities.

Given a many-body fermionic state (or a qubit state which can be mapped to fermions via a JW transformation),  it is interesting to ask how much it differs from a Gaussian state. For example, in the context of many-body physics, quantifying the non-Gaussianity of a state gives us a measure of the interactions of its parent Hamiltonian~\cite{turner2017optimal,jiannis2018quantifying,matos2021emergence, pachos2022quantifying}. In quantum computation theory, instead, fermionic non-Gaussianity can be seen as a resource enabling universal quantum computation in matchgate circuits~\cite{brod2016efficient,hebenstreit2019all}. A natural framework to quantify the deviations of a state from being Gaussian is that of quantum resource theories (QRT)~\cite{chitambar2019quantum}, originally introduced in the context of quantum information theory. 

In a QRT, one indirectly defines a quantity of interest (the resource) in terms of \emph{free} states and operations, which do not display or generate the resource, respectively. The resource of interest (in our case, the fermionic non-Gaussianity) is then quantified by suitable monotones, namely real functions over the Hilbert space that are vanishing on the set of free states and do not increase under free operations~\cite{chitambar2019quantum}. In the QRT of fermionic non-Gaussianity, free states and operations are the Gaussian ones~\cite{bravyi2004lagrangian,bravyi2019simulation,melo2013power,vershynina2014complete,park2021quantifying,cudby2023gaussian,reardon2024improved,dias2024classical,lyu2024fermionic,sierant2026fermionic}.

In this work, we focus on the relative entropy of fermionic non-Gaussianity~\cite{genoni2008quantifying,genoni2010quantifying,marian2013relative,lumia2024measurement, Aditya2025growth}, a popular monotone in the non-Gaussianity QRT, and relate it to the number entropy of the symmetrized state~\cite{klich2008scaling,song2012bipartite,kiefer2020bounds,kiefer2020evidence,kiefer2021slow}. That is, the classical Shannon entropy of the probability distribution of the particle number, which for pure states coincides with the particle-number asymmetry~\cite{vaccaro2008tradeoff}.  Specifically, we derive a lower bound on the relative entropy of non-Gaussianity in terms of the exponential of the Shannon entropy, and study numerically its tightness for large system sizes.  Our bound is non-trivial for large values of the asymmetry and relies on the concentration of the particle-number distribution of (mixed) fermionic Gaussian states.

The implications of our work are two-fold. On the one hand, when the particle-number is experimentally accessible, the particle-number Shannon entropy can be estimated efficiently~\cite{lukin2019probing}, so that our results can be used as a practical way to lower bound non-Gaussianity. In this respect, it is worth mentioning that the total particle number can be measured efficiently also in quantum-circuit setups~\cite{buhrman2024state,piroli2024approximating,rethinasamy2024logarithmic,zi2025shallow}, making our results relevant even for digital quantum platforms. On the other hand, our work provides a link between two different QRTs. Indeed, the notion of asymmetry can also be formalized via a QRT~\cite{gour2008resource,brs-07,gour2009measuring,chitambar2019quantum} (see also Refs.~\cite{marvian2014extending,ares2023entanglement,casini2020entropic, casini2021entropic, magan2021, benedetti2024,summer2025resource} for similar ideas developed in the context of many-body physics and quantum field theory). Our results can then be put in the framework introduced in Ref.~\cite{deneris2025analyzing}, which studied how the free states of one QRT can be resourceful when analyzed through the lenses of a second one. More generally, our findings highlight a non-trivial interplay between non-Gaussianity and particle-number asymmetry.

The rest of this work is organized as follows. We begin in Sec.~\ref{sec:preliminaries}, where we briefly recall the notions of Gaussian states, non-Gaussianity, and asymmetry. In Sec.~\ref{sec:particle_number_gaussian}, we start characterizing Gaussian states in terms of the distribution of the number of particles: after presenting an elementary bound on the particle-number Shannon entropy for Gaussian states in Sec.~\ref{sec:variance_bound}, we give a bound of non-Gaussianity in terms of the minimal trace distance from the set of Gaussian states (also known as interaction distance~\cite{turner2017optimal,jiannis2018quantifying}). Our main results are presented in Sec.~\ref{sec:main_results}. In particular, after some preliminary considerations in Sec.~\ref{sec:preliminary_cond}, in Sec.~\ref{sec:kink} we provide a lower bound on the relative entropy of non-Gaussianity for a special family of states, while a completely general bound is derived in Sec.~\ref{sec:general_bound}. Finally, Sec.~\ref{sec:outlook} contains our conclusions, while a few appendices present the most technical parts of our work.

\section{Preliminaries}
\label{sec:preliminaries}

We begin by recalling some preliminary facts about fermionic Gaussian states, the particle-number Shannon entropy, and the QRT of non-Gaussianity and asymmetry.

\subsection{Fermionic Gaussian states}
We consider a system of $N$ fermionic modes, satisfying canonical anti-commutation relations
\begin{equation}
\begin{split}
     \{c_i, c_j^\dagger\} &= \delta_{i,j}\,,\\
     \{c_i, c_j\} &=\{c_i^\dagger, c_j^\dagger\} = 0\,,
\end{split}
\end{equation}
for $i,j = 1\dots N$. We will denote by $\mathcal{H}$ the corresponding Hilbert space. A mixed, full-rank state $\rho$ on $\mathcal{H}$ is a fermionic Gaussian state if it can be represented as
\begin{equation}\label{eq:exp_quadratic_ham}
    \rho = \frac{1}{Z}e^{-K}\,,
\end{equation}
where $K$ is a quadratic operator of the form
\begin{equation}\label{eq:quadratic_ham}
   \!\!\! K = \sum_{i,j}\left[A_{i,j}c_i^\dagger c_j-A_{i,j}^*c_i c_j^\dagger+B_{i,j}c_i c_j-B_{i,j}^*c_i^\dagger c_j^\dagger\right]\,,
\end{equation}
while $Z = {\rm Tr}[e^{-K}]$ is a normalization constant. Introducing the compact notation
\begin{equation}
\begin{split}
     \mathbf{c} &= (c_1, c_2, \dots c_N, c_1^\dagger, c_2^\dagger, \dots c_N^\dagger)\,,
\end{split}
\end{equation}
the operator $K$ can be rewritten as
\begin{equation}
    K = \mathbf{c}^\dagger M_K \mathbf{c}\,,
\end{equation}
with
\begin{equation}
    M_K = \begin{bmatrix}
A & -B^*\\
B & -A^*
\end{bmatrix}\,.
\end{equation}
More generally, a state (not necessarily mixed) is called Gaussian if it can be obtained from the expression~\eqref{eq:exp_quadratic_ham} by taking the limit to infinity of some of the coefficients $A_{i,j}$ and $B_{i,j}$ in Eq.~\eqref{eq:quadratic_ham}. For instance, a pure Gaussian state can be obtained from a thermal state of the form~\eqref{eq:exp_quadratic_ham}, by taking the limit where the inverse of the temperature is taken to infinty.

We recall that fermionic Gaussian states satisfy Wick's theorem~\cite{bravyi2004lagrangian}. As a result, they are completely characterized by the corresponding two-point correlation functions, which we can collect in the correlation matrix
\begin{equation}
    \Gamma = \begin{bmatrix}
C & F^\dagger\\
F & 1-C^{\rm T}
\end{bmatrix}\,,\label{eq:corr-mat}
\end{equation}
with
\begin{equation}\label{eq:C_and_F}
    C_{i,j} = \tr{[\rho c_i^\dagger c_j]}\quad \text{and} \quad F_{i,j} = \tr{\left[\rho c_i c_j\right]}\,.
\end{equation}
In particular, the correlation matrix $\Gamma$ associated with a Gaussian state $\rho$ gives us access to its von Neumann entropy
\begin{equation}
    S(\rho)=-{\rm Tr}[\rho\log \rho]\,.
\end{equation}
Indeed, one can show that the eigenvalues of $\Gamma$ are real and come in pairs of the form $(\nu_k, 1-\nu_k)$, and that the von Neumann entropy is given by~\cite{vidal2003entanglement, Peschel2003reduced}
\begin{equation}\label{eq:dirac-entropy}
    S(\rho) = -\sum_{k = 1}^{N}\left[\nu_k \log\nu_k + (1-\nu_k)\log(1-\nu_k)\right]\,.
\end{equation}

\subsection{Non-Gaussianity, particle number Shannon entropy, and asymmetry}

In this work, we will focus on the relative entropy of non-Gaussianity~\cite{genoni2008quantifying, genoni2010quantifying,marian2013relative}
\begin{equation}\label{eq:minimization_def}
    {\rm NG}(\rho) ={\rm min}_{\rho'\in\mathcal{G}} S(\rho || \rho')\,,
\end{equation}
where the minimum is taken over $\mathcal{G}$, the set of Gaussian states in $\mathcal{H}$, while we introduced the quantum relative entropy
\begin{equation}
    S(\rho || \sigma)=
    {\rm Tr}[\rho\log \rho]- {\rm Tr}[\rho\log \sigma] \,,
\end{equation}
with the convention that $    S(\rho || \sigma)=\infty$ if ${\rm supp}(\rho)\cap {\rm ker}(\sigma)\neq 0$. While originally introduced in the context of bosonic Gaussian states~\cite{lami2018gaussian}, it is straightforward to see that the relative entropy of non-Gaussianity is a monotone, and hence a good measure of non-Gaussianity, also with respect to the fermionic QRT~\cite{lumia2024measurement}.

As shown in Ref.~\cite{marian2013relative}, given a state $\rho$, the state realizing the minimum in Eq.~\eqref{eq:minimization_def} is its Gaussianification $\rho_G$, namely the Gaussian state with the same correlation matrix (such state always exists). Accordingly, using the properties of $\rho_G$, one can rewrite~\cite{marian2013relative}
\begin{equation}\label{eq:non_gaussianity}
{\rm NG}(\rho) = S(\rho || \rho_G)  = S(\rho_G) -S(\rho)\,.
\end{equation}
Eq.~\eqref{eq:non_gaussianity} does not feature any minimization procedure. In fact, for pure states $S(\rho)=0$, so that ${\rm NG}(\rho) $ can be computed only from the correlation matrix $\Gamma$ as
\begin{equation}
\label{eq:NG_pure_state}
{\rm NG}(\ket{\psi}) = -\sum_{k = 1}^{N}\left[\nu_k \log\nu_k + (1-\nu_k)\log(1-\nu_k)\right],
\end{equation}
where $\{\nu_k,(1-\nu_k)\}$ is the set of eigenvalues of $\Gamma$. In many cases, $\Gamma$ can be computed efficiently even in the many-body setting, justifying the appeal of the relative entropy of non-Gaussianity over different monotones. Note that, for a mixed state $\rho$, the r.h.s of Eq.~\eqref{eq:NG_pure_state} only provides an upper bound for ${\rm NG}(\rho)$.

Next, we discuss the particle-number Shannon entropy. We first introduce the charge
\begin{equation}
 Q=\sum_{j=1}^Nc^\dagger_jc_j\,,
\end{equation}
which is just the particle-number operator. Given a state $\rho$, we can then define the charge probability distribution function
\begin{equation}
    p_q=\tr(\Pi_q\rho),
\end{equation}
where $q=0,1,\ldots N$, while $\Pi_q$ is the projector over the $q$-eigenspace of the charge $Q$. The particle-number Shannon entropy is then
\begin{equation}\label{eq:number_entropy}
    H(\{p_q\})=-\sum_{q=0}^N p_q \log p_q\,.
\end{equation}
For a pure state, the largest the Shannon entropy $H(\{p_q\})$ the more the state breaks the $U(1)$ symmetry associated with the charge $Q$. In fact, the Shannon entropy is closely related to a natural monotone within the QRT of asymmetry~\cite{schuch2004nonlocal,schuch2004quantum, brs-07,vaccaro2008tradeoff,gour2009measuring}, as we now briefly review.

In the asymmetry QRT, one defines free states and operations as those that preserve a certain symmetry group $G$. In our case, this is the $U(1)$ symmetry group associated with the charge $Q$, $\{U=e^{-i\alpha Q}\}_{\alpha}$. The amount of asymmetry in a state $\rho$ is quantified by the $G$-asymmetry~\cite{vaccaro2008tradeoff} (also known as entanglement asymmetry~\cite{ares2023entanglement}) 
\begin{equation}
\label{eq:G-asymmetry_definition}
\Delta S^{U(1)}(\rho)=S(\mathcal{U}[\rho] )-S(\rho )\,,
\end{equation}
where we introduced the twirling operator
    \begin{equation}
    \label{eq: G-twirling definition}
   \mathcal{U}[\rho] := \int_{-\pi}^{\pi} \frac{{\rm d}\alpha}{2\pi}\, e^{-i\alpha Q} \rho\, e^{i\alpha Q}\,.
    \end{equation}
The connection with the particle-number Shannon entropy introduced above comes from the inequality~\cite{mazzoni2026breaking}
\begin{equation}\label{eq:bound_asymm_shanon}
\Delta S^{U(1)}(\rho) \le H(\{p_q\}),
\end{equation}
which is saturated for pure states. Therefore, the particle-number Shannon entropy coincides with the asymmetry for pure states, consistent with physical intuition. 

Recently, the asymmetry has received increasing attention beyond the QRT framework. In particular, the observation that it can be efficiently computed in physically interesting settings led to several studies of asymmetry in the context of quantum thermalization \cite{ares2023entanglement,joshi2024observing, rylands2024microscopic, ares25quantum, bertini2024dynamics, liu2024symmetry, turkeshi2024quantum, ares2025prr}, typical quantum states~\cite{capizzi2024universal, ares2023bh, chen2024ph, russotto25u1, Joshi2026}, QFT \cite{casini2020entropic,casini2021entropic,magan2021,benedetti2024,capizzi2023entanglement, chen2024renyi, fossati2024entanglement, benini2025entanglementCFT, Kusuki2024, fossati2024, fujimura2025entanglement}, and generalized symmetries \cite{benini2025entanglemenHigher, benini25cat, GattoLamas25}. It is also experimentally accessible in several relevant quantum platforms~\cite{joshi2024observing,Xu2025Observation, Joshi2026}.

\section{Particle-number distribution of Gaussian states}
\label{sec:particle_number_gaussian}

In this section, we characterize Gaussian states in terms of the distribution of the number of particles. We first present an elementary bound on the particle-number Shannon entropy for Gaussian states. Then, we show that violations of this bound for a state $\rho$ give us a lower bound on the trace distance of $\rho$ from the set of Gaussian states.

\subsection{A bound on the particle-number Shannon entropy}
\label{sec:variance_bound}

It is well known that Gaussianity constrains the particle-number distribution function. In fact, particle-number fluctuations in Gaussian states have been studied in several settings, including generic states~\cite{klich2014note} and ground states of specific models~\cite{cherng07noise, ivanov13}, as well as during quantum quenches~\cite{groha08fcs, parez21quasiparticle,Parez_2021}.
Previous work has also already unveiled non-trivial connections between the particle-number distribution and other physical quantities. For instance, when the particle number is conserved, its fluctuations within a subsystem were shown to provide a lower bound on the bipartite entanglement entropy~\cite{klich09noise, song2012bipartite, calabrese12exact}. Here, we review some elementary facts yielding a simple upper bound on the particle-number Shannon entropy of Gaussian states.

Given a Gaussian state, we may write the first and second moment of the charge distribution function in terms of the correlation matrices~\eqref{eq:C_and_F}. In particular, a simple application of Wick's theorem yields
\begin{equation}\label{eq:av_charge}
    \braket{Q}=\tr(C),
\end{equation}
and
\begin{equation}
    \braket{Q^2}=\tr(C)+(\tr C)^2-\tr(C^2)+\tr(F^\dagger F).
\end{equation}
Therefore, denoting by $\sigma^2=\langle Q^2\rangle-\langle Q\rangle^2$ the charge variance and using~\eqref{eq:corr-mat}, we have
\begin{equation}\label{eq:sigma_gaussian}
\sigma^2 = 2(\tr(C)-\tr(C^2))+\frac{\tr(\Gamma^2)}{2}-\frac{N}{2}\,.
\end{equation}
Since the eigenvalues of both $\Gamma$ and $C$ are real and lie in the interval $[0, 1]$, it holds $0\leq \tr(C^2)\leq \tr(C)\leq   N$ and $0\leq \tr(\Gamma^2)\leq \tr(\Gamma)=N$.
Applying these inequalities to Eq.~\eqref{eq:sigma_gaussian}, we conclude 
\begin{equation}\label{eq:var_gauss}
 \sigma^2\leq 2N\,.
\end{equation}
In turn, using the known bound for the Shannon entropy of a classical discrete distribution function in terms of its variance~\cite{massey1989entropy, rioul2022gaussian}, we get
\begin{equation}\label{eq:quantitative_bound}
H(\{p_q\})\leq   \frac{1}{2}\log\left[2\pi e\left(2N+\frac{1}{12}\right)\right]\,,
\end{equation}
which is the desired bound on the particle-number Shannon entropy for Gaussian states. Note that, while we focused on the distribution of the eigenvalues of the particle-number operator $Q$, a similar bound could be derived for any local quadratic charge.

Combining with Eq.~\eqref{eq:bound_asymm_shanon}, we also obtain a bound on the asymmetry of a Gaussian state $\rho$,
\begin{equation}\label{eq:bound}
    \Delta S^{U(1)}(\rho)\Big|_{\rho\in \mathcal{G}}\leq \frac{1}{2}\log N[1+o(1)]\,.
\end{equation}
Note that, for a generic state, the asymmetry can be as large as $\log(N+1)$~\cite{gour2009measuring}. Consequently, non-interacting systems cannot maximally break a $U(1)$ symmetry associated to a quadratic charge. In other words, {\it maximal asymmetry requires interactions}.

\subsection{An elementary bound for the trace distance}
\label{sec:trace-distance_bound}

Eq.~\eqref{eq:bound} implies that states with large asymmetry cannot be Gaussian. As an example, one can consider a uniform superposition of charge eigenstates $\ket{v_q}$,
\begin{equation}
	|\psi \rangle =\frac{1}{\sqrt{N+1}}\sum_{q=0}^N |v_{q}\rangle\,.
\end{equation}
The state $\ket{\psi}$ is maximally asymmetric, since $\Delta S^{U(1)}(\ket{\psi})=\log(N+1)$. Thus, regardless of the choice of $\ket{v_q}$, the state is not Gaussian. However, its non-Gaussianity depends on the specific choice of $\ket{v_q}$, and quantifying it is generally non-trivial. The rest of this work is devoted to putting a lower bound on the non-Gaussianity when the particle-number Shannon entropy (and thus, for pure states, the asymmetry) exceeds the r.h.s. of Eq.~\eqref{eq:quantitative_bound}. 

As is turns out, this task is non-trivial when considering the relative entropy of non-Gaussianity, and will be tackled in the next section. Here, we first present a simpler result, giving a lower bound on the trace-distance of a state $\rho$ from the set of Gaussian states. Specifically, denoting as usual the Shannon entropy of $\rho$ by $H(\{p_q\})$, we show
\begin{equation}\label{eq:inequality_trace_distance}
{\rm inf}_{\rho'\in \mathcal{G}}||\rho-\rho'||_1\geq \frac{2 \left(H(\{p_q\}) -\frac{1}{2}\log N + c\right)}{\log N},
\end{equation}
where $c$ is a constant (independent of $N$ and $\rho$), $||A||_1 \equiv \tr\sqrt{A^\dagger A}$, while $\mathcal{G}$ is the set of Gaussian states in $\mathcal{H}$. Note that the l.h.s. is known in the literature as interaction distance~\cite{turner2017optimal,jiannis2018quantifying}.

In order to prove the above inequality, let us consider an arbitrary Gaussian state $\rho'$, and define the probabilities $p_q={\rm Tr}[\rho\Pi_q]$ and $p_q'={\rm Tr}[\rho'\Pi_q]$. From these, we can construct the
mixed states 
\begin{equation}\label{eq:omega_G}
\omega =\sum_{q=0}^N p_q \ket{v_q}\bra{v_q}, \quad
\omega'=\sum_{q=0}^N p'_q\ket{v_q}\bra{v_q},
\end{equation}
which are defined in a Hilbert space of dimension $d=N+1$.
Notice that by construction $S(\omega)=H(\{p_q\})$ and $S(\omega')=H(\{p_q'\})$. Thus, applying the Fannes inequality~\cite{nielsen2010quantum}, we get
\begin{align}
\label{eq:Fannes_ineq}
|H(\{p_q\})-H(\{p_q'\})| &=|S(\omega)-S(\omega')| 
\leq \frac{\log N}{2}||\omega-\omega'||_1+\log 2.
\end{align}

Next, the mixed states in Eq.~\eqref{eq:omega_G} can be written as $\omega = \mathcal{E}(\rho)$, $\omega'=\mathcal{E}(\rho')$, where we introduced the trace preserving channel
\begin{equation}
\mathcal{E}(\cdot)=\sum_{q=0}^N |v_q\rangle\langle v_q | {\rm Tr}[ \Pi_q (\cdot)]\,.
\end{equation}
As a consequence, since the trace distance is non-increasing under CPTP channels~\cite{nielsen2010quantum}, we arrive at
\begin{equation}\label{eq:tr_dist_ineq}
||\omega-\omega'||_1\leq ||\rho-\rho'||_1\,.
\end{equation}
Combining Eqs.~\eqref{eq:Fannes_ineq} and~\eqref{eq:tr_dist_ineq} and taking into account that $H(\{p_q'\})$ is upper bounded by Eq.~\eqref{eq:quantitative_bound}, we find precisely \eqref{eq:inequality_trace_distance}.

\section{A lower bound on fermionic non-Gaussianity}
\label{sec:main_results}

This section contains our main results. After presenting some preliminary considerations in Sec.~\ref{sec:preliminary_cond}, in Sec.~\ref{sec:kink} we provide a lower bound on the relative entropy of non-Gaussianity for a special family of states, while a completely general bound is derived in Sec.~\ref{sec:general_bound}.

\subsection{Preliminary considerations}
\label{sec:preliminary_cond}

Strictly speaking, Eq.~\eqref{eq:inequality_trace_distance} already provides a lower bound on the relative entropy of non-Gaussianity. To see this, we may use Pinsker's inequality \cite{watrous2018theory}, stating that for any two states $\rho$ and $\rho'$,   \begin{equation}
    \left\|\rho-\rho'\right\|_1^2\leq  2	\ln (2) S\left(\rho \| \rho'\right).
\end{equation}
Applying it to $\rho$ and $\rho'=\rho_G$ in Eq.~\eqref{eq:inequality_trace_distance}, we obtain
\begin{equation}\label{eq:pinsker_NG_asymm}
\sqrt{{\rm NG}(\rho)} \geq \frac{2}{\ln 2}\left(\frac{	\Delta S^{U(1)}(\rho) -\frac{1}{2}\log N-c}{\log N}\right).
\end{equation}

Unfortunately, as we will show, the inequality \eqref{eq:pinsker_NG_asymm} provides a bound which is not tight. Before seeing this explicitly, we find it instructive to analyze the asymmetry and the non-Gaussianity of a special family of states, allowing for analytic computations. Below, we show that for this family the relative entropy of non-Gaussianity is bounded by the exponential of the Shannon entropy and grows linearly in $N$ for maximal asymmetry. Conversely, since $\Delta S^{U(1)}(\rho) \le \log(N+1)$, the r.h.s. of Eq.~\eqref{eq:pinsker_NG_asymm} is asymptotically upper-bounded by a constant, suggesting that the bound is indeed not tight. In fact, in Sec.~\ref{sec:general_bound} we will employ a very different strategy and derive a general lower bound on the non-Gaussianity, which recovers a linear growth in $N$ for maximally asymmetric states.

\subsection{Superposition of kink states}\label{sec:kink}

Consider the one-parameter family of states
\begin{equation}\label{eq:kink-family}
    \ket{\psi} = \sum_{k = 0}^{N}\alpha_k\ket{k}\,,\quad \ket{k} = \prod_{j = 1}^{k}c_j^\dagger\ket{0}\,,
\end{equation}
where $\ket{0}$ is the fermionic vacuum annihilated by all $c_j$ and the coefficients are chosen to be
\begin{equation}
\label{eq:coefficients_alpha_kink}
    \alpha_k = \begin{cases}
&\frac{1}{\sqrt{R}}\,, \quad 1\le k \le R\,, \\
&0\,, \quad \text{otherwise}\,,
\end{cases}
\end{equation}
with $0<R\leq N$. Each $\ket{k}$ is a ``kink state'', and is obviously a charge eigenstate with eigenvalue $k$. The simple form of the states~\eqref{eq:kink-family} allows us to perform analytic computations. In addition, by varying the parameter $R$, the asymmetry of $\ket{\psi}$ can be tuned from 0 ($R = 1$) to $\log N$ ($R = N$), allowing us to study the interplay between asymmetry and non-Gaussianity. Quantitatively, by setting $R = \lfloor N^\beta \rfloor$ with $0<\beta \leq1$, the asymmetry of $\ket{\psi}$ reads
\begin{equation}
\label{eq:asymmetry_kink_states_R}
    \Delta S^{U(1)} = -\sum_{k = 1}^{R}|\alpha_k|^2\log|\alpha_k|^2 = \log R \sim \beta\log N\,.
\end{equation}
Note that the state $\ket{\psi}$ breaks the fermion parity, but this is irrelevant for our discussion. In fact, one could also study similar families of states featuring, say, only even kink states, arriving at similar conclusions.

In order to bound the non-Gaussianity of the state~\eqref{eq:kink-family}, we first rewrite Eq.~\eqref{eq:NG_pure_state} as 
\begin{equation}
\label{eq:NG_binary_entropy}
     {\rm NG}(\rho) = \sum_{k = 1}^{N}H_2(\nu_k)\,,
\end{equation}
where $H_2(x)$ is the binary entropy in natural units. Using the simple bound~\cite{topsoe2001bounds}
\begin{equation}
    H_2(x) \geq 4\log (2) \,x(1-x)\,,
\end{equation}
and making use of Eq.~\eqref{eq:corr-mat}, we arrive at the inequality
\begin{equation}\label{eq:tr-bound}
\begin{split}
     {\rm NG}(\rho) 
     \geq 4\log 2\,\tr{\left[C - C^2 - F^\dagger F\right]},
\end{split}
\end{equation}
which is valid for any pure state $\rho$.

Next, as we show in Appendix~\ref{app:kink-computation}, for the family of states ~\eqref{eq:kink-family} we can compute 
\begin{equation}
    \begin{split}
        \tr{\,C} &= \frac{R+1}{2}\,,\\
        \tr{\,C^2} &= \frac{R}{3}+\frac{1}{2}+\frac{1}{6R}\,,\\
        \tr{\,F^\dagger F} &= 2\frac{R-2}{R^2}\,.
    \end{split}
\end{equation}
Finally, recalling that, according to Eq.~\eqref{eq:asymmetry_kink_states_R}, for the states~\eqref{eq:kink-family} $R = e^{\Delta S}$, we arrive at the result
\begin{equation}\label{eq:tr-bound-kink}
    {\rm NG}(\ket{\psi}) \geq 4\log 2\left(4 e^{-2\Delta S} - \frac{13}{6}e^{-\Delta S} + \frac{1}{6}e^{\Delta S}\right)\,.
\end{equation}

\begin{figure}
    \centering
    \includegraphics[width=0.35\textwidth]{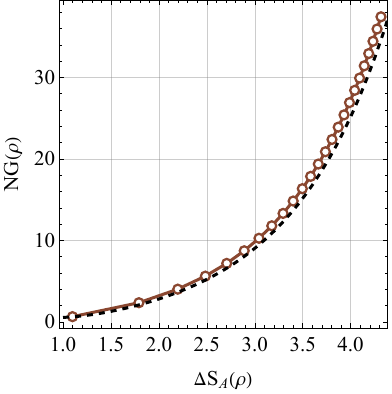}
    \caption{Non-Gaussianity of the superposition of kink states~\eqref{eq:kink-family} plotted against their asymmetry for different values of the parameter $R$ (red) for $N = 100$. The black dashed line denotes the lower bound~\eqref{eq:tr-bound-kink}.}
    \label{fig:kink-numerics}
\end{figure}

The lower bound \eqref{eq:tr-bound-kink} can be verified numerically for all values of $\Delta S^{U(1)}(\ket{\psi})$ by diagonalizing the correlation matrix to obtain the exact value of ${\rm NG}(\ket{\psi})$. We present results for $N = 100$ for all parameters of $R$ (and hence the full range of possible values of the asymmetry) in Fig. \ref{fig:kink-numerics}. Evidently, the lower bound \eqref{eq:tr-bound-kink} follows the actual values very closely and captures the scaling of the non-Gaussianity
\begin{equation}
    \mathrm{NG}(\ket{\psi}) \sim N^\beta + O\left(\frac{1}{N^\beta}\right)\,.
\end{equation}

More generally, we have also studied numerically the non-Gaussianity for different families of states with large asymmetry for small values of $N$. In all cases the bound in Eq.~\eqref{eq:pinsker_NG_asymm} appears to be manifestly non-tight, motivating the derivation of a tighter bound. We do this in Sec.~\ref{sec:general_bound} using a different approach based on a concentration result for the particle-number distribution of (mixed) fermionic Gaussian states.

\subsection{Derivation of the lower-bound }
\label{sec:general_bound}

In this section, we provide a general lower bound on the relative entropy of non-Gaussianity, significantly improving Eq.~\eqref{eq:pinsker_NG_asymm}.  Our strategy is based on the following concentration inequality. Denoting by $\rho$ a Gaussian state with average charge $\mathrm{Tr}[\rho Q]=\bar{q}$, one can prove
\begin{equation}\label{eq:gaussian_inequality_lem}
	P_\rho[|q-\bar{q}|\geq a ]\leq 2\exp\left[ \frac{-a^2 }{2  (\sigma_\rho^2+2a /3)}\right]\,,
\end{equation}
where $\sigma_\rho^2$ is the charge variance of the state $\rho$, while $P_\rho[|q-\bar{q}|\geq a ]$ is the probability that the charge $Q$ takes values outside of the interval $[\bar{q}-a,\bar{q}+a]$. Physically, this result states that the fluctuations of the total charge, $Q$, are strongly suppressed away from their mean. Indeed, for Gaussian states with $\sigma^2\sim O(N)$, fluctuations of size $a \sim \sqrt{N}$ are typical, while extensive deviations, $a\sim N$, are exponentially suppressed in the system size.

While concentration inequalities such as Eq.~\eqref{eq:gaussian_inequality_lem} are well established for pure Gaussian states~\cite{kraus2009pairing,dissertationKraus}, we are not aware of explicit proofs in the case of mixed states. In fact, the extension to mixed states requires a more careful analysis. We prove Eq.~\eqref{eq:gaussian_inequality_lem} in Appendix~\ref{eq:proof}, based on an exact representation of the charge cumulant generating function for (mixed) Gaussian states~\cite{Fagotti_2010,groha08fcs}.

Next, we can summarize the strategy to derive our bound as follows. If a state $\rho$ has a large Shannon entropy, the particle-number distribution function must be anti-concentrated, and therefore different from the one of Gaussian states. In particular, there must be particle-sectors of the Hilbert space where the support of $\rho$ is large, while that of its Gaussianification $\rho_G$ is exponentially small. This difference gives rise to the divergence of the relative entropy between $\rho$ and $\rho_G$ for $N\to\infty$.

To be precise, let us consider a state $\rho$ and denote by $p_q$ the corresponding particle-number probability distribution function. For ease of notation, we will denote
\begin{equation}
    h=e^{H(\{p_q\})}\,,
\end{equation}
and assume $h\geq (N+1)^\gamma$ with $1\geq \gamma\geq 1/2$ (otherwise the bound is trivial). Setting $\varepsilon=P[|q-\bar{q}|>a]$, it is intuitive that if $H(\{p_q\})$ is very large, $\varepsilon$ can not be too small. Introducing $m_a=2\lfloor a\rfloor + 1$, which is the number of integer charge sectors satisfying $|q-\bar{q}|\leq a$, this follows from the Jensen's inequality, yielding
\begin{align}\label{eq:tosaturate}
&H(\{p_q)\}\
\leq \!
H_2(\varepsilon)
+
(1-\varepsilon)\log m_a
+
\varepsilon\log(N+1-m_a)
\leq  \,  
\log 2
+
(1-\varepsilon)\log m_a
+
\varepsilon\log(N+1-m_a)\,,
\end{align}
and therefore
\begin{equation}\label{eq:final_inequality_intermediate}
\!\!P(|q-\bar{q}|>a)
	\geq
	\max\!\left\{
	0,\,
	\frac{H(\{p_q\})\!-\log m_a\!-\!\log 2}
	{\log\!\big(\frac{N+1-m_a}{m_a}\big)}
	\right\}.
\end{equation}
This inequality shows that once the entropy exceeds $\log (2m_a)$, the tails of the probability distribution function must be strictly positive. 

Proceeding with the argument, let $\rho_G$ be the Gaussianification of $\rho$. We define two orthogonal projectors
\begin{align}
	\Pi_A(\alpha)&= \sum_{|q-\bar{q}|\leq c_N(\alpha)}\Pi_q\,,\\
	\Pi_B(\alpha)&= \sum_{|q-\bar{q}|> c_N(\alpha)}\Pi_q\,,
\end{align}
with $c_N(\alpha)=\alpha h$. These operators project onto central and tail charge sectors, respectively. We also introduce the quantum channel
\begin{equation}
	\mathcal{E}_\alpha(\cdot) =	\Pi_A(\alpha) (\cdot ) 	\Pi_A(\alpha)+\Pi_B(\alpha) (\cdot ) 	\Pi_B(\alpha)\,.
\end{equation}
By using the data processing inequality, we get
\begin{equation}\label{eq:non_Gaussianity_rel}
{\rm NG}(\rho)=S(\rho||\rho_G)\geq S(\omega_{\alpha, \rho}||\omega_{\alpha, \rho_G})\,,
\end{equation}
where $\omega_{\alpha, \rho}=\mathcal{E}_\alpha(\rho)$ and similarly for $\rho_G$. Since both states are block-diagonal in this decomposition, we obtain the symmetry-resolved bound
\begin{equation}\label{eq:symm_res_bound}
    S(\omega_{\alpha, \rho}||\omega_{\alpha, \rho_G})\geq p_{B, \rho}\log\frac{p_{B, \rho}}{p_{B, \rho_G}},
\end{equation}
where $p_{B, \rho}=P_{\rho}[|q-\bar{q}|>c_N(\alpha)]$ and $p_{B, \rho_G}=P_{\rho_G}[|q-\bar{q}|>c_N(\alpha)]$. Now, restricting to $\alpha<1/4$, Eq.~\eqref{eq:final_inequality_intermediate} gives us the explicit bound
\begin{equation}\label{eq:sn_alpha}
	p^B_{\alpha,\rho}\geq s_{N}(\alpha)\equiv -\log(4\alpha)\left(\log \frac{(N+1)/h-2\alpha}{2\alpha}\right)^{-1}.
\end{equation}
On the other hand, applying Eq.~\eqref{eq:gaussian_inequality_lem} to the Gaussian state $\rho_G$ with threshold $a=c_N(\alpha)$, we obtain
\begin{equation}\label{eq:temp_in_2}
    p_{B,\rho_G}\leq 2\exp\left[-\frac{c_N(\alpha)^2}{2(\sigma^2_{\rho_G}+2c_N(\alpha)/3)}\right].
\end{equation}
Finally, by plugging Eqs.~\eqref{eq:sn_alpha} and~\eqref{eq:temp_in_2} in Eq.~\eqref{eq:symm_res_bound}, and using Eq.~\eqref{eq:non_Gaussianity_rel}, we arrive at 
\begin{equation}\label{eq:finalbound}
     {\rm NG}(\rho)\geq s_{N}(\alpha)\left[\log \frac{s_{N}(\alpha)}{2}+\frac{c_N(\alpha)^2}{2(\sigma^2_{\rho_G}+2\frac{c_N(\alpha)}{3})}\right]\,,
\end{equation}
where $s_N(\alpha)$ is defined in Eq.~\eqref{eq:sn_alpha}. 

Eq.~\eqref{eq:finalbound} is our main result. While it is true for all values of $\alpha$ with $0<\alpha<1/4$ (so that $s_N(\alpha)>0$), we can choose, for instance, $\alpha=1/8$ to obtain an explicit bound. In this case, the leading scaling behavior of the r.h.s. side can be extracted using that $\sigma^2_{\rho_G}\leq 2N$, yielding
\begin{align}\label{eq:asymptotic_bound}
     {\rm NG}(\rho)&\geq \log(2)(\log [4(N+1)/e^{H(\{p_q\})}-1])^{-1}
    \left[\frac{e^{2H(\{p_q\})}}{256 N}\right](1+o(N))\,,
\end{align}
which displays the previously announced improved scaling over Eq.~\eqref{eq:pinsker_NG_asymm}. For instance, for maximal entropy $H(\{p_q\})\sim \log N$, it yields ${\rm NG}(\rho)> c N (1+o(N))$, where $c$ is a constant, $0<c<1$. This scaling recovers the linear growth of the relative-entropy of non-Gaussianity~\eqref{eq:tr-bound-kink} derived for the family of kink states~\eqref{eq:kink-family}.

For smaller entropy $H(\{p_q\})\sim N^\gamma$ with $\gamma<1$, an additional logarithmic correction appears. In this case, we have not been able to find examples saturating the bound. In fact, for generic states analytic computations are not possible and we are restricted to numerical computations for relatively small system sizes. This makes it difficult to understand whether the scaling predicted by Eq.~\eqref{eq:asymptotic_bound} is tight. We leave this question for future work.

\section{Outlook and discussions}
\label{sec:outlook}

In this work, we have investigated the interplay between two different quantum resources in the many-body setting: non-Gaussianity and asymmetry. In particular, we have derived a general bound on the relative entropy of non-Gaussianity in terms of the Shannon entropy of the particle-number distribution, which coincides with the particle-number asymmetry for pure states. The relevance of our work is two-fold. On the one hand, our results can be viewed as a practical way to lower bound non-Gaussianity. On the other hand, we provide a link between two different QRT in the context of fermionic systems, in the spirit of Ref.~\cite{deneris2025analyzing}.

Our work raises some questions for future research. First, it would be important to understand whether our bound is tight or, conversely, how to improve it. In the case of maximally asymmetric states, we have shown that the scaling of the bound is optimal, but the question remains open for lower asymmetry. 

Second, a natural question pertains to the extension of our results to bosonic systems. In this case, it is easy to see that a bound of the non-Gaussianity in terms of the Shannon entropy as the one we have derived cannot hold. This is because the particle-number Shannon entropy for bosonic Gaussian states is unbounded, as it can be seen for a single bosonic mode. However, one may wonder whether a bound on non-Gaussianity can be obtained by studying refined properties of the particle-number distribution. This would be particularly interesting, given the prominent role that bosonic Gaussian states play in the context of quantum optics~\cite{lami2018gaussian}. We leave these questions for future research.

\prlsection{Acknowledgments} LP acknowledges fruitful discussions with Ludovico Lami, especially regarding possible extensions to the bosonic case. This work was funded by the European Union (ERC, QUANTHEM, 101114881 and ERC, MOSE, 101199196). Views and opinions expressed are however those of the author(s) only and do not necessarily reflect those of the European Union or the European Research Council Executive Agency. Neither the European Union nor the granting authority can be held responsible for them.


\appendix

\section{Correlation matrices and lower bound for the superposition of kink states}\label{app:kink-computation}

Here we provide some additional details about the computation of the lower bound for the non-Gaussianity of the family of states considered in Sec.~\ref{sec:kink}, namely
\begin{equation}
    \ket{\psi} = \sum_{k = 0}^{N}\alpha_k\ket{k}\,,\quad \ket{k} = \prod_{j = 1}^{k}c_j^\dagger\ket{0}\,,
\end{equation}
where the coefficients $\alpha_k$ are:
\begin{equation}
    \alpha_k = \begin{cases}
&\frac{1}{\sqrt{R}}\,, \quad 1\le k \le R \\
&0\,, \quad \text{otherwise}
\end{cases}\quad,
\end{equation}
with $0<R\leq N$.

Our starting point is the inequality \eqref{eq:tr-bound}. For the state we are considering, the matrix elements of $C$ and $F$ are immediately obtained,
\begin{equation}
\begin{split}
     C_{i,j} &= \tr{[\rho c_i^\dagger c_j]} = \delta_{i,j}\sum_{k = i}^{N}|\alpha_k|^2 \, ,\\
     F_{i,j} &= \tr{\left[\rho c_i c_j\right]} = \alpha_{i-2}\alpha_{i}\delta_{i, j+1} - \alpha_{j-2}\alpha_{j}\delta_{i, j-1}\,.
\end{split}
\end{equation}

We now proceed to compute $\tr{\left[C - C^2 - F^\dagger F\right]}$ term by term. Crucially, the only non-trivial part of the computation is keeping track of the summation limits. Thus, noting that $\alpha_k =0$ whenever $k\notin[1,R]$, from the above matrix elements we readily obtain
\begin{equation}
    \tr{[C]} = \sum_{i = 1}^{R}  \frac{1}{R}(R-i+1)
    = \frac{R+1}{2}\,.
\end{equation}
For the second term, $\tr[C^2]$, the matrix elements are given by
\begin{equation}
    (C^2)_{i,j} = \delta_{i,j} \frac{1}{R^2}(R-i+1)^2\quad\text{for}\quad 1\leq i\leq R
\end{equation}
and so
\begin{equation}
    \tr{[C^2]} = \sum_{i = 1}^{R}\frac{1}{R^2}(R-i+1)^2 = \frac{R}{3}+\frac{1}{2}+\frac{1}{6R}\,.
\end{equation}
For the third term, $\tr[F^\dagger F]$, we write
\begin{equation}
    (F^\dagger F)_{i,i} = \sum_{k = 1}^{R}\left(\alpha_{i}\alpha_{i-2}\alpha_{i}\alpha_{i-2}\delta_{k,i-1}+\alpha_{k}\alpha_{k-2}\alpha_{k}\alpha_{k-2}\delta_{i,k-1}\right) = \alpha_{i}\alpha_{i-2}\alpha_{i}\alpha_{i-2}+\alpha_{i+1}\alpha_{i-1}\alpha_{i+1}\alpha_{i-1}\,,
\end{equation}
yielding
\begin{equation}
\tr{\left[ F^\dagger F\right]} = \sum_{i = 1}^{R}(\alpha_{i}\alpha_{i-2}\alpha_{i}\alpha_{i-2}+\alpha_{i+1}\alpha_{i-1}\alpha_{i+1}\alpha_{i-1}) =\frac{2(R-2)}{R^2}\,.
\end{equation}
Combining the above results, we finally have
\begin{equation}
    \tr{\left[C - C^2 - F^\dagger F\right]} = 4\frac{1}{R^2} - \frac{13}{6}\frac{1}{R} + \frac{1}{6}R\, ,
\end{equation}
which leads to Eq.~\eqref{eq:tr-bound-kink} by substituting $R=e^{H(\{p_q\})}=e^{\Delta S}$.

\section{Proof of the concentration inequality for mixed states}\label{eq:proof}

In this Appendix, we prove the concentration result~\eqref{eq:gaussian_inequality_lem}. To this end, we first define the probability distribution $P_{\rho}[q]$ associated to a generic state $\rho$ as
\begin{equation}
    P_{\rho}[q] = \tr{[\rho\Pi_q]}\,,
\end{equation}
where $\Pi_q$ is the projector onto the $q$-eigenspace of the charge operator $Q$. The moment-generating function of such probability distribution is defined
\begin{equation}
    \chi_\rho(t) = \mathbb{E}[e^{tQ}] = \sum_{q}P_{\rho}[q]e^{tq} = \tr{[\rho e^{tQ}]}\,.
\end{equation}
If $\rho$ is a Gaussian state, then $\chi_\rho(t)$ can be computed explicitly by the formula~\cite{Fagotti_2010,groha08fcs}
\begin{equation}\label{eq:gaussian-generating-func}
    \chi_\rho(t) = (2\cosh(t/2))^N\sqrt{\textrm{det}\left(\frac{\mathbf{1}-\tanh{(\frac{t}{2})\Gamma_M \Omega}}{2}\right)}\,.
\end{equation}
Here we introduced the correlation matrix $\Gamma_M$ in the majorana basis, which is defined by
\begin{equation}
    \Gamma_M = \frac{i}{2}\tr{\left(\rho [\gamma_i, \gamma_j]\right)}\, ,
\end{equation}
where
\begin{equation}
\begin{split}
     &\gamma_{2 j -1} = c_j^\dagger + c_j\\
     &\gamma_{2 j} = i(c_j^\dagger - c_j)\,.
\end{split}
\end{equation}
The matrix $\Omega$ is a $2N \times 2N$ real matrix whose only non-zero elements are
\begin{equation}
\begin{split}
    \Omega_{2j-1,2j} &= -1\\
\Omega_{2j,2j-1} &=1\,.
\end{split}
\end{equation}
The matrix $\Gamma_M \Omega$ is not necessarily diagonalizable. Still, Eq.~\eqref{eq:gaussian-generating-func} is well-defined since $\Gamma_M \Omega$ can be characterized by the eigenvalues appearing in the Jordan form. The cumulant generating function can be expanded
\begin{equation}
\begin{split}
       \log \chi_\rho(t)&=N \log (\cosh t/2)+\frac{1}{2}{\rm Tr}\left\{\log [\mathbf{1}-\tanh(t/2) \Gamma_M \Omega]\right\}\nonumber\\
	&=\frac{1}{2}{\rm Tr}\left\{\log [\cosh(t/2)\mathbf{1}-\sinh(t/2) \Gamma_M \Omega]\right\}=N \log (\cosh t/2)-\frac{1}{2}\sum_{n=1}^\infty\frac{\tanh^n(t/2) }{n}{\rm Tr}[(\Gamma_M \Omega)^n]\,.\label{eq:log_generating_function} 
\end{split}
\end{equation}
This series is convergent for all $t\in\mathbb{R}$ since $\tr[(\Gamma_M\Omega)^n]\leq 2N$ due to $||\Gamma_M\Omega||_\infty \leq 1$. Indeed, the first two terms in the series give us the first two cumulants, namely the mean and variance of the charge $Q$:
\begin{align}
    \braket{Q}&=\frac{i}{2}\sum_{j=1}^N\braket{\gamma_{2j-1}\gamma_{2j}}=-\frac{1}{4} {\rm Tr}[\Gamma_M \Omega]\,,\\
    \braket{Q^2}- \braket{Q}^2&=\frac{N}{4}-\frac{1}{8} {\rm Tr}[\Gamma_M \Omega\Gamma_M \Omega]\,.
\end{align}

The matrix $\Gamma_M\Omega$ is a product of two skew-symmetric matrices, one of which is non-degenerate. As such, its spectrum is doubly degenerate \cite{Fagotti_2010, ikramov2009product}. Denoting as $\{\mu_j\}_{j=1}^N$ the non-doubled eigenvalues of $\Gamma_M\Omega$, we can rewrite the cumulant generating function \eqref{eq:gaussian-generating-func} as
\begin{equation}\label{eq:gaussian-generating-func-2}
    \log\chi(t) =\log \mathbb{E} [e^{Xt}]
= \sum_{j=1}^N \log\left[\cosh(t/2)-\mu_j\sinh(t/2)\right]\,,
\end{equation}
from which it is apparent that the cumulants can be also expressed in terms the eigenvalues $\mu_j$. In particular, for the first two we have
\begin{align}
	\braket{Q}&=-\frac{1}{2}\sum_{j=1}^N\mu_j\,,\\
	\braket{Q^2}- \braket{Q}^2&= \frac{1}{4}\left[\sum_{j=1}^N (1-\mu_j^2)\right]\,.
\end{align}
Importantly, the eigenvalues of $\Gamma_M\Omega$ are not necessarily real. It is however instructive to look at the case where all $\mu_j$ are real. In this case, since the eigenvalues satisfy $|\mu_j|\leq 1$, it is obvious that Eq.~\eqref{eq:gaussian-generating-func-2} can be interpreted as the cumulant generating function of the sum of $N$ independent random variables. Explicitly, for each $j$ we define a $\{\pm 1\}$--valued random variable $Z_j$ by
\begin{equation}
P(Z_j = 1) = \frac{1-\mu_j}{2},
\qquad
P(Z_j = -1) = \frac{1+\mu_j}{2}\,,
\end{equation}
where $P(Z_j = a)$ denotes the probability that $Z_j=a$. It is then immediate to see that
\begin{equation}
\mathbb{E}[e^{s Z_j}]
= \cosh s - \mu_j \sinh s\,.
\end{equation}
Taking $s = t/2$ then leads to
\begin{equation}\label{eq:generating_z}
\log \chi(t)
= \sum_{j=1}^N \log \mathbb{E} [e^{(t/2) Z_j}]\,,
\end{equation}
which is the cumulant generating function of $\frac12 \sum_{j=1}^N Z_j$.

In general, however, the $\mu_j$ are complex, and so the above treatment must be refined. This is easy to do using some additional structure of the eigenvalues $\mu_j$. Indeed, we note that $\Gamma_M \Omega$ is a $2N\times 2N$ real matrix. For any real matrix, the complex eigenvalues come in pairs $\mu_j=\nu_j$, $\mu_{j+1}=\bar{\nu}_j$ (with the same algebraic and geometric multiplicity). For such a pair, we can then define a valid real random variable $Y_j$,  corresponding to the sum $(Z_j+Z_{j+1})/2$, as we describe below. 

Formally, suppose we order $\{\mu_j\}_{j=1}^{N}=\{\xi_j\}_{j=1}^{N-K}\cup\{\nu_j\}_{j=N-K+1}^{N-K/2}\cup\{\bar{\nu}_j\}_{j=N-K/2+1}^{N}$, where $\xi_j$ are the real eigenvalues, while $\{\nu_j\}_{j=N-K+1}^{N-K/2}$ is the set obtained by selecting one of the two conjugate eigenvalues in each pair. For each $j\in[N-K+1,N-K/2]$, define a random variable $Y_j\in [-1,0,1]$ with the corresponding probabilities given by
\begin{align}
	P(Y_j = 1)& = \frac{1-\nu_j}{2}\frac{1-\bar{\nu}_j}{2},\\
	P(Y_j = 0)& = \frac{1-\nu_j}{2}\frac{1+\bar{\nu}_j}{2}+ \frac{1+\nu_j}{2}\frac{1-\bar{\nu}_j}{2}\,,\\
	P(Y_j = -1) &=\frac{1+\nu_j}{2}\frac{1+\bar{\nu}_j}{2}\,.
\end{align}
Because $|\nu_j|\leq 1$ (which follows from $||\Gamma_M\Omega||_{\infty}\leq 1)$, the probabilities $	P(Y_j = a)$, with $a=1,0,-1$, are positive, while we can verify
\begin{equation}
\sum_{a=-1}^1P(Y_j = a)=1\,.
\end{equation} 
Therefore, $\{P(Y_j = a)\}_j$ is a legitimate probability distribution function. 
For $j \in [1, N-K]$, corresponding to the real eigenvalues $\xi_j$, we define
$Y_{j}\in [-1/2, 1/2]$, with probability distribution function
\begin{equation}
P(Y_j = 1/2) = \frac{1-\mu_j}{2},
\qquad
P(Y_j = -1/2) = \frac{1+\mu_j}{2}.
\end{equation}
Because $\mu_j$ are real, $\{P(Y_j = a)\}_j$ is again a legitimate probability distribution function. 

Now, following the previous steps, we have
\begin{equation}\label{eq:generating_y}
	\log \chi(t)
	= \sum_{j=1}^{N-K/2} \log \mathbb{E} [e^{t Y_j}],
\end{equation}
which is the cumulant generating function of 
\begin{equation}
Y= \sum_{j=1}^{N-K/2} Y_j. 
\end{equation}
Note that, contrary to Eq.~\eqref{eq:generating_z}, in Eq.~\eqref{eq:generating_y} there is no factor $1/2$ in the exponent, consistent with the fact that $\sum_{j} Y_j=(1/2)\sum_j Z_j$.

As a result, the cumulants (and moments) of the random variable $Y$ coincide with the cumulants (and moments) of the classical variable associated to the probability distribution $P_\rho[q]={\rm Tr}[\rho \Pi_q]$. Therefore, we can write
\begin{equation}
	P_Y[q]=P_\rho[q]\,.
\end{equation}
This result is important, because it tells us that $P_\rho[q]$ coincides with the probability distribution function of the sum of $N-K/2$ random variables, for which concentration inequalities are well known.
In particular, using that the independent variables are bounded $|Y_j|\leq1$, we use Bernstein's inequality \cite{vershynin20218high}:
\begin{equation}
	P\left(\left|\sum_{i=1}^{N-K/2} (Y_i-\mathbb{E}[Y_i])\right| \geq a\right) \leq 2\exp \left(-\frac{\frac{1}{2} a^2}{\sigma^2+\frac{2}{3} a}\right)\,
\end{equation}
where
\begin{equation}
\sigma^2=\mathbb{E}[Y^2]-\mathbb{E}[Y]^2= \braket{Q^2}- \braket{Q}^2\,,
\end{equation}
which concludes the proof of the concentration inequality \eqref{eq:gaussian_inequality_lem}.

\bibliography{bibliography}

\end{document}